# A unified theory for perfect absorption in ultra-thin absorptive films with constant tangential electric or magnetic fields

Jie Luo, Sucheng Li, Bo Hou, Yun Lai*

College of Physics, Optoelectronics and Energy & Collaborative Innovation Center of Suzhou Nano Science and Technology, Soochow University, Suzhou 215006, China

*laiyun@suda.edu.cn

The maximal absorption rate of ultra-thin films is 50% under the condition that the tangential electric (or magnetic) field is almost constant across the film in symmetrical environment. However, with certain reflectors, the absorption rate can be greatly increased, to even perfect absorption (100%). In this work, we explicitly derive the general conditions of the ultra-thin absorptive film parameters to achieve perfect absorption with general types of reflectors under the condition that the tangential electric (or magnetic) field is almost constant across the film. We find that the parameters of the film can be classified into three groups, exhibiting: 1) a large permittivity (permeability), 2) a near-zero permeability (permittivity), or 3) a suitable combination of the permittivity and the permeability, respectively. Interestingly, the latter two cases demonstrate extraordinary absorption in ultra-thin films with almost vanishing losses. Our work serves as a guide for designing ultra-thin perfect absorbers with general types of reflectors.

# I. INTRODUCTION

Lossy materials can absorb electromagnetic (EM) wave energy, but if the impedance is mismatched with air, EM waves will be reflected at the surface, leading to inefficiency in energy absorption. Traditional methods to minimize such a reflection include anti-reflection films with a quarter-wavelength thickness [1], wavelength-scale-thick films [2-5], micro-structures with gradually varying impedance [6], etc. Recently, new methods to achieve good absorbers have been proposed in the fields of metamaterials and plasmonics, which provide highly controllable EM response in different frequency bands by artificial EM structures [7-19]. Comparing with traditional absorbers, many interesting characteristics such as broadband, wide-angle and polarization-independent behaviors have been demonstrated. Especially, coherent perfect absorbers with coherent illumination have also been proposed to achieve perfect absorption (PA), i.e. 100% absorption [20-24], which can be regarded as the reverse process of laser generation.

Among so many types of absorbers, ultra-thin absorptive films are an especially interesting class of absorbers, with an extraordinarily thin, light and simple system. With the thickness much smaller than the wavelength, the phase change of waves inside the film is often negligible, and usually either the tangential electric or magnetic field is almost constant across the ultra-thin film. Various types of ultra-thin absorbers have been studied in various frequency regimes, including conductive films in microwave and terahertz (THz) regimes [1, 23-25], metal in optical regimes [23], organic materials [26], semi-conductors [27-32], graphene [33, 34], magnetic materials [35, 36], epsilon(mu)-near-zero media [37-43], metamaterial absorber with strong magnetic responses [44], etc. It is known that for such ultra-thin films, the absorption rate is up to 50% in symmetrical environment [1, 33, 45, 46]. However, such a limit can be broken by introducing symmetrical coherent illumination [15, 16] or specific reflectors. Various types of ultra-thin films with different reflectors have been proposed and demonstrated, which show that absorption rate >50% is possible. However, a unified theory for ultra-thin absorbers with general reflectors is still lacking.

In this paper, we analytically derive the general solutions of the film parameters for PA

for a general class of reflectors characterized by reflected waves with an amplitude $\eta_E$ ($0 < \eta_E \leq 1$) and a phase shift $\phi_E$ ($0 \leq \phi_E < 2\pi$). Our work gives a unified theory for realizing PA by using ultra-thin films with constant tangential electric or magnetic fields under both normal and oblique incidence.

When propagating EM waves are absorbed by an ultra-thin homogeneous film, the cases may be divided into three classes. First, the tangential electric and magnetic fields are both almost constant across the film. In this case, the transmission is still unity and there would be no absorption inside the film. Second, both the tangential electric and magnetic fields are non-constant across the film. This indicates that the wavelength inside the film is comparable to the film thickness. The mechanisms of such ultra-short wavelengths include meta-surface and high-impedance surfaces [47-49], standing wave resonances [17] and surface plasmon resonances [18, 19], etc. It is rather complicated to draw a unified theory for this case. Third, only the tangential electric field or magnetic field is almost constant across the absorber film. Interestingly, in this case, it is possible to obtain clear and complete conditions of the film for the PA. Another advantage of this case over the second one is the extreme broad bandwidth of PA in low frequencies, as we shall demonstrate here. In this paper, we mainly focus on the third case with constant tangential electric fields. For the third case with constant tangential magnetic fields, similar results can be obtained (see the formula for the case of constant tangential magnetic fields in the Appendix).

Specifically, for the cases of constant tangential electric (magnetic) field across the film, we find out unified solutions to the required permittivity and permeability of the absorber films, showing that the general solutions of the film parameters can be classified into three groups, in which PA is mainly induced by 1) a large the permittivity (permeability), 2) a near zero permeability (permittivity), or 3) a suitable combination of the permittivity and permeability, respectively. Especially, the third case corresponds to PA at large incident angles with near-unity permittivity and permeability. Interestingly, although the second and third cases originate in different physical mechanisms, they both demonstrate extraordinary absorption in arbitrarily thin films with almost vanishing loss. Our work serves as a guide to design the ultra-thin film and reflectors for PA.

## II. GENERAL SOLUTIONS TO THE PERFECT ABSORBPTION

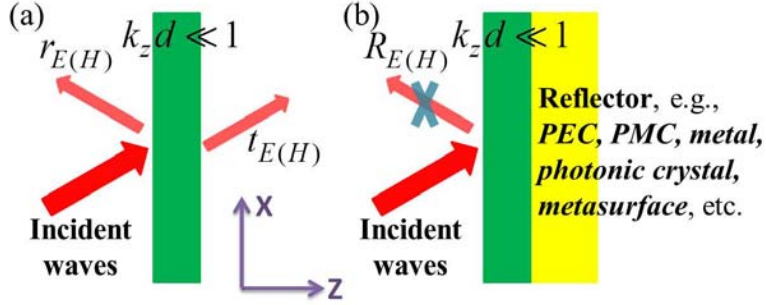

FIG. 1. Schematic graph of the transmission and reflection of EM waves on an ultra-thin absorptive film (a) without and (b) with a reflector attached behind the film.

The system of our investigation is illustrated in Fig. 1(a). In this work, we consider the case in which the film is much thinner compared to the wavelength inside the film, and the phase change across the film is negligible, i.e.

$$k_z^{TE} d = k_0 d \sqrt{\frac{\mu_x}{\mu_z}\left(\varepsilon_y \mu_z - \sin^2\theta\right)} \ll 1 \qquad \text{1(a)},$$

$$k_z^{TM} d = k_0 d \sqrt{\frac{\varepsilon_x}{\varepsilon_z}\left(\mu_y \varepsilon_z - \sin^2\theta\right)} \ll 1 \qquad \text{1(b)},$$

for transverse electric (TE) polarization ($E_y$) and transverse magnetic (TM) polarization ($H_y$), respectively. Here, $k_z^{TE}$ and $k_z^{TM}$ are the $z$ component of wave vectors for TE and TM polarizations, respectively. $d$, $\theta$, and $k_0$ are the thickness of the absorber film, the incident angle, and the wave number in free space, respectively.

In such an ultra-thin limit, usually either the tangential electric or magnetic field is almost constant across the film. This leads to

$$1 + r_{E(H)} \approx t_{E(H)}, \qquad (2)$$

where $r_{E(H)}$ and $t_{E(H)}$ are the reflection and transmission coefficients of the film defined on the tangential electric (magnetic) field [50] as illustrated in Fig. 1(a).

If both the tangential electric and magnetic fields are almost constant across the film, then the transmission is almost unity, and the ultra-thin film does not absorb energy. If only the tangential electric or magnetic field is almost constant across the film in symmetrical environment, there exists a maximal absorption of 50% which corresponds to $-r_{E(H)} \approx t_{E(H)} \approx 0.5$ [1, 33, 45, 46]. If both the electric and magnetic fields are not constant across the film, then the wavelength inside the film is usually comparable to the film thickness. In this case, there is no 50% absorption limit but there would be complicated resonances.

In this paper, we will focus on the case in which only the tangential electric or magnetic field is almost constant across the film. For simplicity, we consider the case that only the electric field is almost constant across the film, i.e., $1 + r_E \approx t_E$ and $1 + r_H \neq t_H$. The case of constant tangential magnetic field can be obtained similarly (see the details in the Appendix). By using the transfer matrix method [50] and Eq. (1), we obtain the reflection and transmission coefficients through an ultra-thin film characterized by relative permittivity (permeability) tensor $\bar{\varepsilon}(\bar{\mu}) = \begin{pmatrix} \varepsilon(\mu)_x & 0 & 0 \\ 0 & \varepsilon(\mu)_y & 0 \\ 0 & 0 & \varepsilon(\mu)_z \end{pmatrix}$ as,

$$t_E = \frac{2}{2 - i\left(f_1^{TE} + f_2^{TE}\right)x^{TE}}, \quad r_E = -\frac{i}{2}t_E\left(f_1^{TE} - f_2^{TE}\right)x^{TE},$$

and $\quad t_H = \frac{2}{2 - i\left(f_1^{TE} + f_2^{TE}\right)x^{TE}}, \quad r_H = \frac{i}{2}t_H\left(f_1^{TE} - f_2^{TE}\right)x^{TE}$ (3a)

for TE polarizations, and

$$t_E = \frac{2}{2 - i\left(f_1^{TM} + f_2^{TM}\right)x^{TM}}, \quad r_E = \frac{i}{2}t_E\left(f_1^{TM} - f_2^{TM}\right)x^{TM},$$

and $\quad t_H = \frac{2}{2 - i\left(f_1^{TM} + f_2^{TM}\right)x^{TM}}, \quad r_H = -\frac{i}{2}t_H\left(f_1^{TM} - f_2^{TM}\right)x^{TM}$ (3b)

for TM polarizations, where $f_1^{TE} = \cos\theta\sqrt{\frac{\mu_x\mu_z}{\varepsilon_y\mu_z - \sin^2\theta}}$, $f_2^{TE} = \frac{1}{\cos\theta}\sqrt{\frac{\varepsilon_y\mu_z - \sin^2\theta}{\mu_x\mu_z}}$,

$$f_1^{TM} = \cos\theta \sqrt{\frac{\varepsilon_x \varepsilon_z}{\mu_y \varepsilon_z - \sin^2\theta}}, \quad \text{and} \quad f_2^{TM} = \frac{1}{\cos\theta}\sqrt{\frac{\mu_y \varepsilon_z - \sin^2\theta}{\varepsilon_x \varepsilon_z}}, \quad x^{TE} = k_z^{TE} d = k_0 d \sqrt{\frac{\mu_x}{\mu_z}\left(\varepsilon_y \mu_z - \sin^2\theta\right)},$$

and $x^{TM} = k_z^{TM} d = k_0 d \sqrt{\frac{\varepsilon_x}{\varepsilon_z}\left(\mu_y \varepsilon_z - \sin^2\theta\right)}$.

By inserting Eq. (3) into $1 + r_E \approx t_E$ and $1 + r_H \neq t_H$, we get

$$f_1^{TE} x^{TE} = \cos\theta |\mu_x| k_0 d \approx 0, \quad f_2^{TE} x^{TE} = \frac{1}{\cos\theta}\left|\frac{\varepsilon_y \mu_z - \sin^2\theta}{\mu_z}\right| k_0 d \neq 0 \tag{4a}$$

for TE polarizations, and

$$f_2^{TM} x^{TM} = \frac{1}{\cos\theta}\left|\frac{\mu_y \varepsilon_z - \sin^2\theta}{\varepsilon_z}\right| k_0 d \approx 0, \quad f_1^{TM} x^{TM} = \cos\theta |\varepsilon_x| k_0 d \neq 0 \tag{4b}$$

for TM polarizations.

Equation (4) presents the necessary conditions of the film parameters to have almost constant electric field but a variant magnetic field across the film. Under normal incidence, the conditions reduce to $|\mu_x| k_0 d \approx 0$ and $|\varepsilon_y| k_0 d \neq 0$ for TE polarizations, and $|\mu_y| k_0 d \approx 0$ and $|\varepsilon_x| k_0 d \neq 0$ for TM polarizations. Obviously, such cases satisfy the ultra-thin condition of Eq. (1), which reduces to $k_0 d \sqrt{\mu_x \varepsilon_y} \ll 1$ and $k_0 d \sqrt{\varepsilon_x \mu_y} \ll 1$ in the case of normal incidence.

To break the maximal absorption limit of 50% and achieve PA for such a film, we attach the film with a reflector, as shown in Fig. 1(b). A reflector can be any materials with zero transmission rate, including perfect electric conductor (PEC), perfect magnetic conductor (PMC), lossy metal, photonic crystals within band gaps, reflective high impedance surfaces or meta-surfaces, dielectric medium with total internal reflection, layered materials, etc. With reflectors attached, the absorption rate is only associated with reflection rate. Here, we characterize the properties of the reflector by using the amplitude $\eta_E$ ($0 < \eta_E \leq 1$) and phase shift $\phi_E$ ($0 \leq \phi_E < 2\pi$) of reflected waves. By using Eq. (2), the total reflection coefficient is obtained as,

$$\begin{aligned} R_E &= (t_E - 1) + t_E \eta_E e^{i\phi_E} t_E + t_E \eta_E e^{i\phi_E}(t_E - 1)\eta_E e^{i\phi_E} t_E + t_E \eta_E e^{i\phi_E}(t_E - 1)\eta_E e^{i\phi_E}(t_E - 1)\eta_E e^{i\phi_E} t_E + \ldots \\ &= \frac{\left(\eta_E^{-1} e^{-i\phi_E} + 2\right) t_E - \eta_E^{-1} e^{-i\phi_E} - 1}{\eta_E^{-1} e^{-i\phi_E} + 1 - t_E}. \end{aligned} \tag{5}$$

The condition of PA is $R_E = 0$. By letting $R_E = 0$ and considering Eq. (2), we have

$$t_E = \frac{\eta_E^{-1} e^{-i\phi_E} + 1}{\eta_E^{-1} e^{-i\phi_E} + 2} \quad \text{and} \quad r_E = \frac{-1}{\eta_E^{-1} e^{-i\phi_E} + 2}. \tag{6}$$

By inserting Eq. (3) and Eq. (4) into Eq. (6), we analytically obtain the parameter conditions of the ultra-thin film to achieve PA with a reflector of $(\eta_E, \phi_E)$,

$$\frac{\sin^2 \theta}{\mu_z} - \varepsilon_y + i \frac{2\cos\theta}{k_0 d \left(\eta_E^{-1} e^{-i\phi_E} + 1\right)} = 0 \tag{7a}$$

for TE polarizations, and

$$-\varepsilon_x + i \frac{2}{k_0 d \left(\eta_E^{-1} e^{-i\phi_E} + 1\right) \cos\theta} = 0 \tag{7b}$$

for TM polarizations.

Equation (7) is the unified solution to the required permittivity and permeability of the absorber film for the case with constant tangential electric fields, showing many interesting properties for the ultra-thin film to achieve PA. For TE polarizations, based on Eq. (7), we can classify the solutions into three groups. Case 1). $\left|\frac{\sin^2\theta}{\mu_z}\right| \ll |\varepsilon_y| \approx \left|\frac{2\cos\theta}{k_0 d \left(\eta_E^{-1} e^{-i\phi_E} + 1\right)}\right|$. In this case, a large permittivity of $\varepsilon_y = i \frac{2\cos\theta}{k_0 d \left(\eta_E^{-1} e^{-i\phi_E} + 1\right)}$ is required to achieve PA. And the change of permeability in a range would not affect the PA. Case 2). $|\varepsilon_y| \ll \left|\frac{\sin^2\theta}{\mu_z}\right| \approx \left|\frac{2\cos\theta}{k_0 d \left(\eta_E^{-1} e^{-i\phi_E} + 1\right)}\right|$. In this case, a small permeability of $\mu_z = \frac{i}{2} k_0 d \left(\eta_E^{-1} e^{-i\phi_E} + 1\right) \sin\theta \tan\theta$ is required to achieve PA, while the permittivity can vary in a range without affecting the PA. Case 3). The terms in Eq. 7(a) $\left|\frac{\sin^2\theta}{\mu_z}\right|$ and $|\varepsilon_y|$ are comparable. In this case, PA can be achieved for a suitable combination of the permittivity and permeability.

For TM polarizations, PA can only be achieved by using a large permittivity $\varepsilon_x = i \frac{2}{k_0 d \left(\eta_E^{-1} e^{-i\phi_E} + 1\right) \cos\theta}$, and is totally independent of the permeability.

Another information that we can retrieve from Eq. (7) is that when $\eta_E^{-1} e^{-i\phi_E} + 1 = 0$, Eq. (7)

can never be satisfied. $\eta_E^{-1} e^{-i\phi_E} + 1 = 0$ leads to $\eta_E = 1$, $\phi_E = \pi$, implying a PEC reflector. Physically, a PEC reflector forces the tangential electric field to be zero. As a result, the tangential electric field will be zero throughout the film, making PA impossible. As a result, the PEC reflector cannot be applied in the case of constant tangential electric fields.

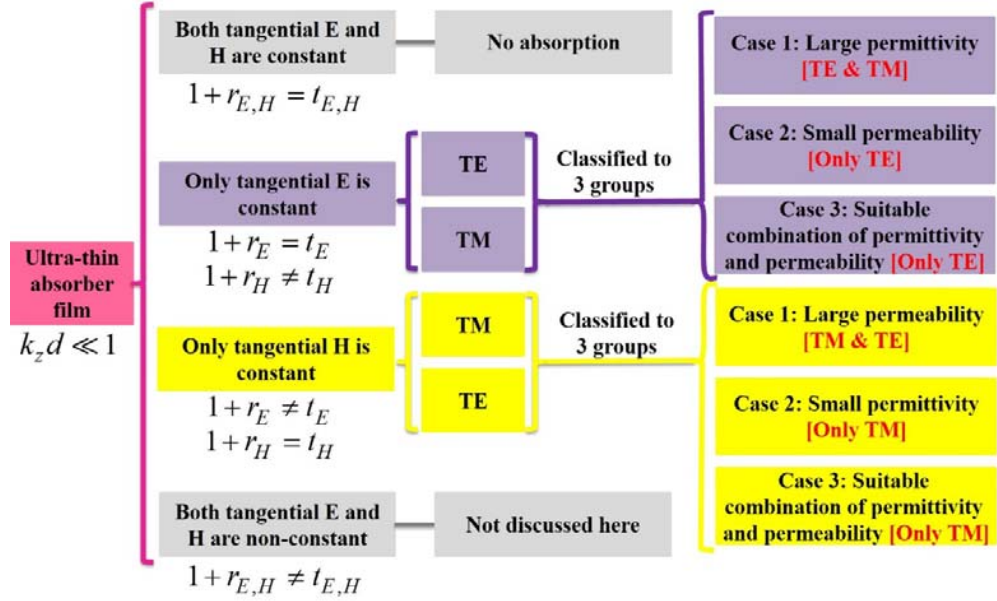

FIG. 2. Outline of the unified theory for the ultra-thin absorber with constant tangential electric or magnetic fields.

The outline of the unified theory is illustrated in Fig. 2, showing different groups of solutions for the ultra-thin absorber film with either electric or magnetic fields. In the following sections, we will discuss in details the three groups of solutions for constant tangential electric field across the film, in which PA is due to large permittivity, small permeability, and a suitable combination of the permittivity and permeability, respectively. For the case of constant tangential magnetic field, the previous equations simply need to exchange the permittivity with permeability, which is reported in the Appendix.

### III. CASE 1: PERFECT ABSORPTION DUE TO LARGE PERMITTIVITY

In case 1, PA for TM polarizations is mainly dependent on the value of $\varepsilon_x$. On the other hand, for TE polarizations, when $|\mu_z| \gg \left| \frac{k_0 d \left( \eta_E^{-1} e^{-i\phi_E} + 1 \right)}{2\cos\theta} \sin^2\theta \right|$, the condition of PA is reduced to,

$$\varepsilon_y = i \frac{2\cos\theta}{k_0 d \left( \eta_E^{-1} e^{-i\phi_E} + 1 \right)} = \frac{2\cos\theta}{k_0 d} \left( -\frac{\eta_E^{-1} \sin\phi_E}{\eta_E^{-2} + 2\eta_E^{-1} \cos\phi_E + 1} + i \frac{\eta_E^{-1} \cos\phi_E + 1}{\eta_E^{-2} + 2\eta_E^{-1} \cos\phi_E + 1} \right). \tag{8}$$

Under normal incidence with $\theta \to 0$, $|\mu_z| \gg \left| \frac{k_0 d \left( \eta_E^{-1} e^{-i\phi_E} + 1 \right)}{2\cos\theta} \sin^2\theta \right|$ is always fulfilled. While under oblique incidence, $|\mu_z| \gg \left| \frac{k_0 d \left( \eta_E^{-1} e^{-i\phi_E} + 1 \right)}{2\cos\theta} \sin^2\theta \right|$ is also satisfied due to ultra-thin thickness $k_0 d \to 0$, except for the cases of $\theta \to \pm 90°$ or $|\mu_z| \to 0$.

### A. PMC reflector and coherent illumination

In particular, if the reflector is a PMC with $\eta_E = 1$ and $\phi_E = 0$, the required relative permittivities will be simplified to,

$$\varepsilon_y = i \frac{\cos\theta}{k_0 d} \quad \text{and} \quad \varepsilon_x = i \frac{1}{k_0 d \cos\theta} \tag{9}$$

for TE and TM polarized waves, respectively. This indicates that pure imaginary permittivities are required for PA with a PMC reflector. Interestingly, metals in low frequency regime naturally exhibit almost pure imaginary permittivity. For a metal with a conductivity $\sigma_0$, the relative permittivity is described as $\varepsilon_r = 1 + i \frac{\sigma(\omega)}{k_0} Z_0 \approx i \frac{\sigma_0}{k_0} Z_0$, where $Z_0$ is the impedance of free space. Therefore, we can see that both the dispersion of required permittivity in Eq. (9) and the material dispersion of the metal are proportional to $i/\omega$. Interestingly, we can tune the thickness $d$ or the conductivity $\sigma_0$ of the conductive film to obtain the dispersion match. And it is easily found that the sheet resistance, $R_s$, defined as $1/(\sigma_0 d)$, is required to be

$$R_s = 1/(\sigma_0 d) = Z_0/\cos\theta \quad \text{and} \quad R_s = 1/(\sigma_0 d) = Z_0 \cos\theta \tag{10}$$

for TE and TM polarizations, respectively. Clearly, the required $R_s$ is a function of the incident angle only rather than the operating frequency or film thickness. Such a frequency-independent property indicates that ultra-thin film absorbers may be applied to achieve ultra-broadband absorbers, for all frequencies below 100GHz, including radio waves and microwaves. Even in THz and optical regime, some conductive materials e.g., inconel [25], tungsten [23], graphene [33, 34, 51], approximately possess permittivities proportional to $i/\omega$, and can still be exploited to achieve broadband absorption.

To verify the analytical result, we carry out numerical simulations based on the finite element software, COMSOL Multi-physics, as shown in Fig. 3. Figure 3(a) presents the normalized amplitude of total electric field $|\mathbf{E}|/|\mathbf{E}_{in}|$ (blue solid lines) and magnetic field $|\mathbf{H}|/|\mathbf{H}_{in}|$ (red dashed lines) under normal incidence on an ultra-thin conductive film with a PMC reflector. The absorber film on the PMC reflector is characterized by a thickness of $d = \lambda_0/100$ and an isotropic relative permittivity of $\varepsilon_r = 15.92i$. The unit field amplitude in the free space region indicates there is no reflection and proves the PA. In Fig. 3(a), it is also seen that the electric field is almost constant inside the absorber film, while the amplitude of magnetic field is rapidly decaying in the film. In Fig. 3(b), we plot the absorptance as a function of the incidence angle. Despite that the PA condition is broken for oblique incidence, it is seen that high absorptance rate >0.9 is observed in a wide angle range. By applying the transfer matrix method, we find that both the absorptance for TE and TM polarized waves for the ultra-thin absorber film with $R_s = Z_0$ can be written as

$$A = 1 - \tan^4(\theta/2), \tag{11}$$

which is confirmed by numerical simulation as shown in Fig. 3(b).

Although we have shown above that the ultra-thin film with suitable resistance and a PMC reflector can realize PA. However, the realization of PMC reflector itself is another important issue. Recently, high impedance surfaces or meta-surfaces have also been extensively investigated and they can work as effective PMCs [52-55]. These structures are deep sub-wavelength, and support engineering of the amplitude $\eta_E$ and phase change $\phi_E$ of reflected waves.

In most cases, although $\phi_E$ can be tuned to be zero, but $\eta_E$ would be smaller than unity due to the losses of the reflector. Thus, the required permittivity of the absorber film would be changed, and the real part of the permittivity may occur as the change of $\phi_E$. As the result, the conductive film with almost pure imaginary permittivity may fail to perfectly absorb the EM waves. We would like to look into the issue and study in which range of $\eta_E$ and $\phi_E$, the conductive film, with the real part of the permittivity is much smaller than its imaginary part, can be applied for PA.

In Figs. 3(c) and 3(d) we plot the required sheet resistance $R_s$ and the ratio $|\text{Re}(\varepsilon_r)/\text{Im}(\varepsilon_r)|$ as the function of $\eta_E$ and $\phi_E$, respectively. It is seen that the influence of phase shift $\phi_E$ on $R_s$ is weak while decreased $\eta_E$ leads to increased $R_s$. However, the real part $\text{Re}(\varepsilon_r)$ increases rapidly with increased $\phi_E$, but is insensitive to changes in $\eta_E$. Therefore, in practical designs, we should pay more attention on the tuning of phase change $\phi_E$ when we use a conductive film with pure imaginary permittivity as the absorber film.

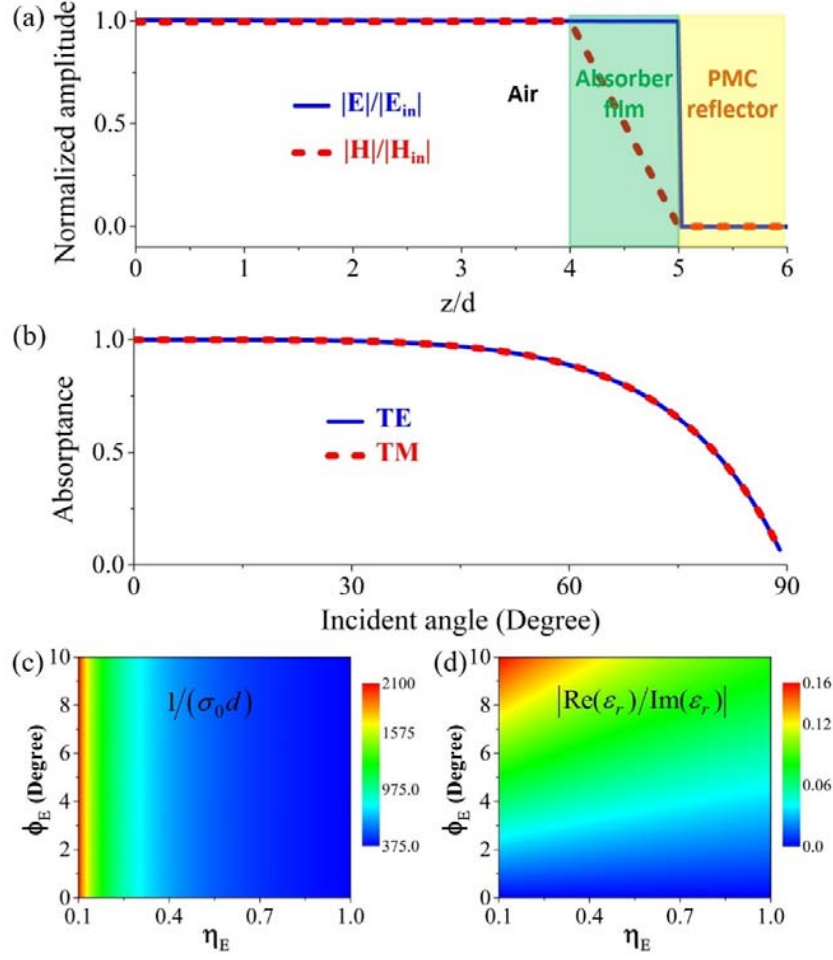

FIG. 3. (a) The normalized amplitude electric field $|\mathbf{E}|/|\mathbf{E}_{in}|$ (blue solid lines) and magnetic field $|\mathbf{H}|/|\mathbf{H}_{in}|$ (red dashed lines) in the case of PA on an ultra-thin film characterized by $\varepsilon_r = 15.92i$, $\mu_r = 1$ and $d = \lambda_0/100$ with a PMC reflector under normal incidence. (b) Absorptance as a function of the incident angle for both TE (black solid lines) and TM (red dashed lines) polarizations. (c) The required sheet resistance $1/(\sigma_0 d)$ for the PA and (d) the ratio $|\text{Re}(\varepsilon_r)/\text{Im}(\varepsilon_r)|$ as the function of the $\eta_E$ and $\phi_E$ under normal incidence.

Besides high impedance surfaces and meta-surfaces, other methods for mimicking PMC include PEC coated with a high-index dielectric layer with optical thickness $\lambda_0/4$ [1, 25, 34, 45, 56] (known as the Salisbury screen) or an anisotropic layer through transformation optics [57], dielectric Bragg mirror with a spacer layer [24], epsilon-near-zero media with dielectric defects [58], or dielectric resonators [59], etc. However, these methods require a finite thickness of the wavelength order.

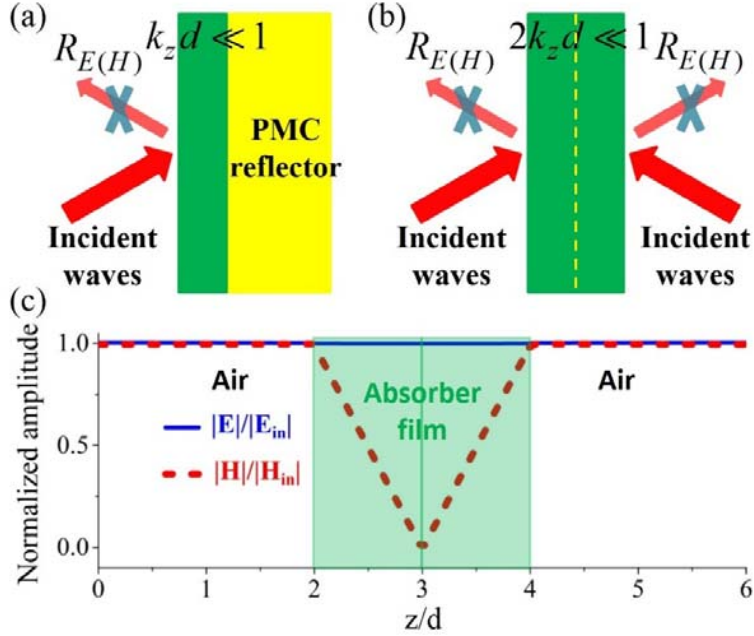

FIG. 4. Schematic graphs of (a) an ultra-thin absorber film with a PMC reflector attached, (b) an equivalent double-layered absorber film with waves symmetrically incident from both sides. (c) The normalized amplitude electric field $|\mathbf{E}|/|\mathbf{E}_{in}|$ (blue solid lines) and magnetic field $|\mathbf{H}|/|\mathbf{H}_{in}|$ (red dashed lines) under normal incidence for symmetrical illumination. The absorber film is characterized by $\varepsilon_r = 15.92i$, $\mu_r = 1$ and $d = \lambda_0/50$.

Although conductive films can almost exactly satisfy the dispersion relation to obtain the PA, in practice, the PMC reflector can be realized only in a narrow frequency [53]. Consequently, the bandwidth of PA lies in the bandwidth of the PMC reflector. Actually, a PMC reflector is equivalent to a special mirror. The absorber film with a PMC reflector is equivalent to a double-layered absorber film with symmetrical coherent illumination, as illustrated in Figs. 4(a) and 4(b). According to Eq. (3), the reflection and transmission coefficients of the double-layered film are $r_E = -0.5$ and $t_E = 0.5$. Thus, the total reflection coefficient $R_E = r_E + t_E = 0$. Figure 4(c) shows the normalized amplitude distribution in the double-layered absorber film with a total thickness of $d = \lambda_0/50$. The relative permittivity and relative permeability are chosen as $\varepsilon_r = 15.92i$ and $\mu_r = 1$, respectively, which are the same as those in Fig. 2(a). The incoming waves normally incident from both side in a symmetrical manner. It is seen from Fig. 4(c) that the normalized amplitude of electric field in both air and absorber regions is nearly unity, indicating almost PA. However, the magnetic field amplitude exhibits a dip at the center. In such a way, the bandwidth of PMC reflector can be avoided and ultra-broadband absorption is indeed possible. Po et al. have theoretically shown that a 17nm tungsten film can absorb most of the EM waves with wavelength ranging from 800nm to 1500nm [23]. Recently, we have experimentally demonstrated the PA in the frequency range from 6GHz to 18GHz by using ultra-thin conductive films in the microwave regime [24]. It is verified that the PA is indeed independent of frequency for a suitable sheet resistance.

On the other hand, to obtain broadband PA in the case of constant tangential magnetic fields, a large permeability proportional to $i/\omega$ is required, which has been pointed out in Refs. [35, 36].

## B. Other phase-controllable and absorptive reflectors

In most practical cases, the reflector has a reflection amplitude less than unity and a reflection shift $\phi_E$ substantially different from 0 and $\pi$. In high frequency regime, such as infrared and optical frequencies, the PMC reflectors and absorber films with large pure imaginary part are difficult to achieve. Reflectors such as dielectric Bragg mirror [26], meta-surfaces [7], metal [30-32] and sapphire [27-29] which can induce a reflection shift $\phi_E$ substantially different from 0 and $\pi$ and result in a complex value of $\eta_E^{-1} e^{-i\phi_E}$, have also been investigated. As a consequence, the required permittivity is also a complex number as described by Eqs. (7b) and (8), instead of a pure imaginary number as in the case of PMC reflectors.

Particularly, under normal incidence, the required permittivities turn out to be,

$$\varepsilon_x = \varepsilon_y = \frac{2}{k_0 d}\left(-\frac{\eta_E^{-1}\sin\phi_E}{\eta_E^{-2}+2\eta_E^{-1}\cos\phi_E+1}+i\frac{\eta_E^{-1}\cos\phi_E+1}{\eta_E^{-2}+2\eta_E^{-1}\cos\phi_E+1}\right) \quad (12)$$

Here, we demonstrate a specific example which was described in Ref. [31]. Suppose that the working wavelength is $\lambda_0 = 532 nm$ and the reflector is selected as gold characterized by a refractive index of $n_{Gold} = 0.44 + 2.24i$. The thickness of the absorber film is fixed as $d = 10 nm$. Through Eq. (12), we can easily calculate the required refractive index of the absorber film for PA as $n = \sqrt{\varepsilon_x} = \sqrt{\varepsilon_y} = 4.39 + 0.54i$, which is quite near the value $n = 4.3 + 0.71i$ found by Kats et al. [31] through numerical search. Our unified theory can provide a simple way to obtain required parameters of ultra-thin film and reflector for PA. It is worth noting that experimental realization of such ultra-thin absorbers can be based on conductive organic materials [26], semiconductors [27, 30, 31], conductive oxides [28], etc, in optical and infrared range.

## IV. CASE 2: PERFECT ABSORPTION DUE TO SMALL PERMEABILITY

Here, we consider a situation with $|\varepsilon_y| \ll \left|\frac{\sin^2\theta}{\mu_z}\right| \approx \left|\frac{2\cos\theta}{k_0 d\left(\eta_E^{-1}e^{-i\phi_E}+1\right)}\right|$ (see Eq. (7)) for the case of continuous tangential electric field. In this case, the condition of the PA for TE polarized waves is

$$\mu_z = \frac{i}{2}k_0 d\left(\eta_E^{-1}e^{-i\phi_E}+1\right)\sin\theta\tan\theta, \quad (13)$$

and is less dependent on the permittivity of the absorber film. If the reflector is a PMC, we have,

$$\text{Re}(\mu_z) \ll \text{Im}(\mu_z) \to 0 \text{ and } \frac{d}{\lambda_0} = \frac{\text{Im}(\mu_z)}{2\pi\sin\theta\tan\theta}. \quad (14)$$

On the other hand, if the tangential magnetic field is almost constant inside the absorber film, we get $\text{Re}(\varepsilon_z) \ll \text{Im}(\varepsilon_z) \to 0$ and $\frac{d}{\lambda_0} = \frac{\text{Im}(\varepsilon_z)}{2\pi\sin\theta\tan\theta}$ for TM polarizations (see the formula for the case of constant tangential magnetic fields in the Appendix), which have been theoretically demonstrated by Harbecke et al. [37], Feng et al. [39] and Badsha et al. [43], and experimentally verified by Zhong et al. [41] by using metamaterials and Luk et al. [42] by using indium tin oxide just above the plasma frequency. In practice, several methods to obtain near-zero permittivity or permeability have been proposed, which can be found in Ref. [42-48].

Although such PA is only achievable for oblique incidence [38-41], and generally works for narrow band due to intrinsically dispersive zero-index materials [60-66], one interesting advantage of such PA is that there exists a linear relationship between the thickness and the loss, which means that the thickness of the absorber can be pushed to zero by reducing the material loss to zero. As a result, an arbitrarily thin perfect absorber with near-zero value parameters is possible [38-41], which is different from common understanding that thinner absorber film needs a larger absorption part to maintain the same absorption.

To clarify the physical origin of the extraordinary absorption, we numerically study an ultra-thin film characterized by $\mu_z = 0.01814i$, $\mu_x = \varepsilon = 1$ and $d = \lambda_0/100$ with a PMC reflector attached. TE polarized waves are incident from air under an incident angle of $\theta = 30 \deg$. The normalized amplitude distribution in Fig. 5 confirms the homogeneous electric fields $E_y$ (blue solid lines) and the decaying x-component of magnetic field $H_x$ (red dashed lines) inside the film. It is also seen that the z-component magnetic field $H_z$ (purple dotted lines) is enhanced by about 55 times in the absorber film, which is required by the continuity of $B_z$ at the interface. $H_z$ increases rapidly as the decrease of the thickness of the absorber film, leading to the extraordinary absorption in the ultra-thin film.

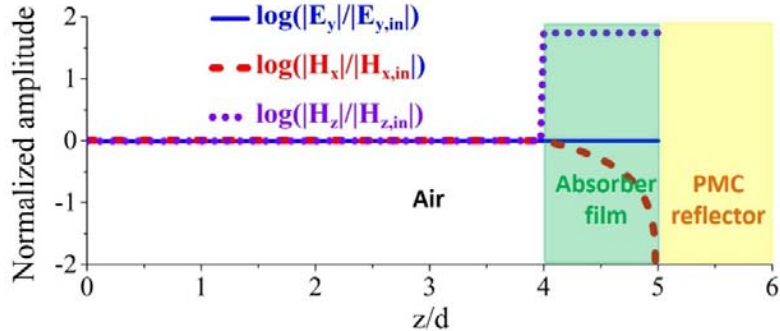

FIG. 5. The normalized amplitude electric field $\log(|E_y|/|E_{y,in}|)$ (blue solid lines),

x-component of magnetic field $\log(|H_x|/|H_{x,in}|)$ (red dashed lines) and z-component of magnetic field $\log(|H_z|/|H_{z,in}|)$ (purple dotted lines) for EM waves of TE polarization incident on an absorber film characterized by $\mu_z = 0.01814i$, $\mu_x = \varepsilon = 1$ and $d = \lambda_0/100$ with a PMC reflector under the incidence angle of $\theta = 30\deg$.

## V. CASE 3: PERFECT ABSORPTION DUE TO A SUITABLE COMBINATION OF PERMITTIVITY AND PERMEABILITY

Now we consider the case that PA is induced by a suitable combination of the permittivity and permeability of the absorber film when $\left|\frac{\sin^2\theta}{\mu_z}\right|$ and $|\varepsilon_y|$ are comparable, i.e. $\left|\frac{2\cos\theta}{k_0 d(\eta_E^{-1} e^{-i\phi_E} + 1)}\right| \ll \left|\frac{\sin^2\theta}{\mu_z}\right| \approx |\varepsilon_y|$ or $|\varepsilon_y| \sim \left|\frac{\sin^2\theta}{\mu_z}\right| \sim \left|\frac{2\cos\theta}{k_0 d(\eta_E^{-1} e^{-i\phi_E} + 1)}\right|$. For a film with a fixed $k_0 d$ and under oblique incidence, when $\theta \to \pm 90\deg$ or $\eta_E \to 0$, we have $\left|\frac{2\cos\theta}{k_0 d(\eta_E^{-1} e^{-i\phi_E} + 1)}\right| \to 0$ and the PA would depend on a suitable choice of $\mu_z$ and $\varepsilon_y$. $\eta_E \to 0$ indicates that the reflector alone can absorb a lot of the incident waves and there is less need for the film. However, when $\theta \to \pm 90\deg$, we find the condition of PA turns into $\frac{1}{\mu_z} - \varepsilon_y \approx 0$. That is, as the increase of the incident angle, the real part of $\varepsilon_y \mu_z$ tends to be unity, while its imaginary part tends to zero. This indicates that a vanishing loss is capable of achieving PA for the case of very large incident angle, similar to the extraordinary absorption in the near zero permeability film in the case 2. However, the physical origin is totally different. The extraordinary high absorption in the zero-index media is caused by the greatly enhanced fields in the longitudinal direction. While here, it is seen that when $\theta \to \pm 90\deg$, we have $\sqrt{\mu_z \varepsilon_y} \approx 1$, which means that the absorber film has a refractive index close to that of air, as illustrated by Fig. 6(a). So the waves propagate almost along the $x$ direction in phase with the waves in air, and therefore have an extremely long absorption distance. Eventually, the waves are gradually absorbed with a tiny loss.

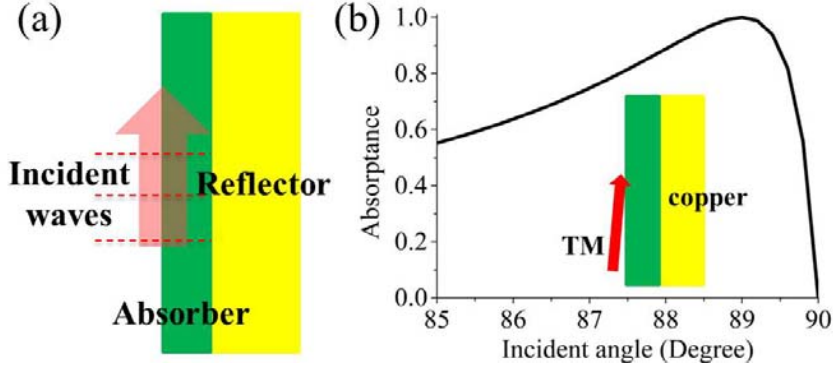

FIG. 6. (a) Schematic graph of PA when waves are incident on the ultra-thin absorptive film with extremely large angles, i.e. $\theta \to \pm 90 \deg$. The waves in the ultra-thin film and air are in phase. (b) The absorptance as the function of the incident angle for an absorber film with $\varepsilon = 0.9278 + 0.2565i$, $\mu = 1$ and $d = \lambda_0/100$. The reflector is chosen as copper with DC conductivity being $\sigma_0 = 6.5 \times 10^7 \Omega^{-1} m^{-1}$.

In most practical cases, the situation $|\varepsilon_y| \sim \left|\frac{\sin^2\theta}{\mu_z}\right| \sim \left|\frac{2\cos\theta}{k_0 d\left(\eta_E^{-1} e^{-i\phi_E} + 1\right)}\right|$ is satisfied under oblique incidence. Here we take an example of constant tangential magnetic field, which requires $\left|\frac{\sin^2\theta}{\varepsilon_z}\right| \sim |\mu_y| \sim \left|\frac{2\cos\theta}{k_0 d\left(\eta_H^{-1} e^{-i\phi_H} + 1\right)}\right|$ for TM polarized waves. We exploit practical copper as the reflector, whose DC conductivity is about $\sigma_0 = 6.5 \times 10^7 \Omega^{-1} m^{-1}$, the required relative permeability of the absorber film with $\mu = 1$ and $d = \lambda_0/100$ calculated from Eq. (7) is $\varepsilon = 0.9278 + 0.2565i$ under an incident angle of $\theta = 89 \deg$. Moreover, we plot the absorptance with respect to the incident angle in Fig. 6(b), showing that the absorption is indeed sensitive to the variation of incident angle.

## VI. CONCLUSION

In conclusion, we have explicitly derived the general solutions to the PA for an ultra-thin absorptive film with a general reflector. The cases of constant tangential electric field are investigated in detail. We find the solutions for TE polarizations can be classified to three groups in which PA is induced by large permittivity, small permeability or a suitable combination of permittivity and permeability. The solutions for TM polarizations can only be induced by large permittivity for the case of constant tangential electric fields. Based on our theory, we have not only physically explained results in previous literature, but also found new mechanisms to achieve PA. Especially, we find that ultra-thin conductive films with a PMC reflector can achieve frequency-independent PA, which has the same principle as the PA achieved under symmetrical coherent illumination. We also find a new type of PA mechanism for large incident angle (case 3), which might be useful to achieve PA in waveguide structures. Although the cases in this paper were demonstrated for constant tangential electric field, similar analyses would easily be applied to the cases of constant tangential magnetic field due to the symmetry nature of electric and magnetic fields in the Maxwell's equations.

Finally, we note that in this work we have only discussed the cases in which the ultra-thin films are composed of be homogeneous materials or effective media, of artificial structures. However, in practical designs, we can directly apply Eq. (6) to find out the solutions to the PA by adjusting the reflection and transmission coefficients. For reflectors, our theory applies to the cases with no diffractions, in which only the 0-th order reflection coefficients should be taken into account.

**ACKNOWLEDGEMENTS**

This work is supported by the State Key Program for Basic Research of China (No. 2012CB921501), National Natural Science Foundation of China (No. 11104196, No.11104198, No. 11374224), Natural Science Foundation of Jiangsu Province (No. BK2011277), Program for New Century Excellent Talents in University (NCET), and a Project Funded by the Priority Academic Program Development of Jiangsu Higher Education Institutions (PAPD).

**APPENDIX**

## A. Formula for the case of constant tangential electric fields

Parameter conditions of the ultra-thin film with constant tangential electric fields to achieve PA with a reflector of $(\eta_E, \phi_E)$ (cannot be PEC),

$$\frac{\sin^2\theta}{\mu_z} - \varepsilon_y + i\frac{2\cos\theta}{k_0 d\left(\eta_E^{-1} e^{-i\phi_E} + 1\right)} = 0 \tag{A1a}$$

for TE polarizations, and

$$-\varepsilon_x + i\frac{2}{k_0 d\left(\eta_E^{-1} e^{-i\phi_E} + 1\right)\cos\theta} = 0 \tag{A1b}$$

for TM polarizations.

The solutions in Eq. (A1) can be classified to 3 groups:

### 1. Perfect absorption due to large permittivity

This case occurs under $\left|\frac{\sin^2\theta}{\mu_z}\right| \ll |\varepsilon_y| \approx \left|\frac{2\cos\theta}{k_0 d\left(\eta_E^{-1} e^{-i\phi_E} + 1\right)}\right|$ for TE polarizations. And for TM polarizations, due to $\frac{1}{k_0 d} \gg 1$, the required $\varepsilon_x$ should be large. Thus, Eq. (A1) can be simplified to,

$$\varepsilon_y = i\frac{2\cos\theta}{k_0 d\left(\eta_E^{-1} e^{-i\phi_E} + 1\right)} \implies \varepsilon_y = i\frac{\cos\theta}{k_0 d} \quad \text{(PMC reflector)} \tag{A2a}$$

for TE polarizations, and

$$\varepsilon_x = i\frac{2}{k_0 d\left(\eta_E^{-1} e^{-i\phi_E} + 1\right)\cos\theta} \implies \varepsilon_x = i\frac{1}{k_0 d \cos\theta} \quad \text{(PMC reflector)} \tag{A2b}$$

for TM polarizations.

### 2. Perfect absorption due to small permeability

This situation occurs only for TE polarizations when $|\varepsilon_y| \ll \left|\frac{\sin^2\theta}{\mu_z}\right| \approx \left|\frac{2\cos\theta}{k_0 d\left(\eta_E^{-1} e^{-i\phi_E} + 1\right)}\right|$, and Eq. (A1) can be simplified to,

$$\mu_z = \frac{i}{2}k_0 d\left(\eta_E^{-1} e^{-i\phi_E} + 1\right)\sin\theta\tan\theta$$

$$\implies \text{Re}(\mu_z) \ll \text{Im}(\mu_z) \to 0 \text{ and } \frac{d}{\lambda_0} = \frac{\text{Im}(\mu_z)}{2\pi\sin\theta\tan\theta} \quad \text{(PMC reflector)} \tag{A3}$$

for TE polarizations.

### 3. Perfect absorption due to a suitable combination of permittivity and permeability

This case occurs only for TE polarizations when $\left|\dfrac{2\cos\theta}{k_0 d\left(\eta_E^{-1}e^{-i\phi_E}+1\right)}\right| \ll \left|\dfrac{\sin^2\theta}{\mu_z}\right| \approx |\varepsilon_y|$ or

$|\varepsilon_y| \sim \left|\dfrac{\sin^2\theta}{\mu_z}\right| \sim \left|\dfrac{2\cos\theta}{k_0 d\left(\eta_E^{-1}e^{-i\phi_E}+1\right)}\right|$. In particular, if $\theta \to \pm 90\deg$ or $\eta_E \to 0$, we have

$$\sqrt{\mu_z \varepsilon_y} \approx 1 \tag{A4}$$

for TE polarizations.

### B. Formula for the case of constant tangential magnetic fields

Parameter conditions of the ultra-thin film with constant tangential magnetic fields to achieve PA with a reflector of $(\eta_H, \phi_H)$ (cannot be PMC),

$$\dfrac{\sin^2\theta}{\varepsilon_z} - \mu_y + i\dfrac{2\cos\theta}{k_0 d\left(\eta_H^{-1}e^{-i\phi_H}+1\right)} = 0 \tag{A5a}$$

for TM polarizations, and

$$-\mu_x + i\dfrac{2}{k_0 d\left(\eta_H^{-1}e^{-i\phi_H}+1\right)\cos\theta} = 0 \tag{A5b}$$

for TE polarizations.

The solutions in Eq. (A5) can be classified to 3 groups:

### 1. Perfect absorption due to large permeability

This case occurs under $\left|\dfrac{\sin^2\theta}{\varepsilon_z}\right| \ll |\mu_y| \approx \left|\dfrac{2\cos\theta}{k_0 d\left(\eta_H^{-1}e^{-i\phi_H}+1\right)}\right|$ for TM polarizations. And for TE polarizations, due to $\dfrac{1}{k_0 d} \gg 1$, the required $\mu_x$ should be large. Thus, Eq. (A5) can be simplified to,

$$\mu_y = i\dfrac{2\cos\theta}{k_0 d\left(\eta_H^{-1}e^{-i\phi_H}+1\right)} \implies \mu_y = i\dfrac{\cos\theta}{k_0 d} \quad \text{(PEC reflector)} \tag{A6a}$$

for TM polarizations, and

$$\mu_x = i\dfrac{2}{k_0 d\left(\eta_H^{-1}e^{-i\phi_H}+1\right)\cos\theta} \implies \mu_x = i\dfrac{1}{k_0 d \cos\theta} \quad \text{(PEC reflector)} \tag{A6b}$$

for TE polarizations.

### 2. Perfect absorption due to small permittivity

This situation occurs only for TM polarizations when $|\mu_y| \ll \left|\frac{\sin^2\theta}{\varepsilon_z}\right| \approx \left|\frac{2\cos\theta}{k_0 d\left(\eta_H^{-1} e^{-i\phi_H} + 1\right)}\right|$, and

Eq. (A5) can be simplified to,

$$\varepsilon_z = \frac{i}{2} k_0 d \left(\eta_H^{-1} e^{-i\phi_H} + 1\right) \sin\theta \tan\theta$$

$$\Rightarrow \quad \text{Re}(\varepsilon_z) \ll \text{Im}(\varepsilon_z) \to 0 \quad \text{and} \quad \frac{d}{\lambda_0} = \frac{\text{Im}(\varepsilon_z)}{2\pi \sin\theta \tan\theta} \quad \text{(PEC reflector)} \quad (A7)$$

for TM polarizations.

### *3. Perfect absorption due to a suitable combination of permittivity and permeability*

This case occurs only for TM polarizations when $\left|\frac{2\cos\theta}{k_0 d\left(\eta_H^{-1} e^{-i\phi_H} + 1\right)}\right| \ll \left|\frac{\sin^2\theta}{\varepsilon_z}\right| \approx |\mu_y|$ or $|\mu_y| \sim \left|\frac{\sin^2\theta}{\varepsilon_z}\right| \sim \left|\frac{2\cos\theta}{k_0 d\left(\eta_H^{-1} e^{-i\phi_H} + 1\right)}\right|$. In particular, if $\theta \to \pm 90\deg$ or $\eta_H \to 0$, we have

$$\sqrt{\varepsilon_z \mu_y} \approx 1 \tag{A8}$$

for TM polarizations.